\documentclass[runningheads]{llncs}
\usepackage{graphicx}
\usepackage{booktabs}
\usepackage{multirow}
\usepackage{caption}
\usepackage{bbding}
\usepackage{cite}
\usepackage{hyperref}

\begin{document}
\title{ContriMix: Scalable stain color augmentation for domain generalization without domain labels in digital pathology}

\author{%
Tan H. Nguyen$^{1}$ \quad Dinkar Juyal$^{1}$ \quad Jin Li$^1$ \quad Aaditya Prakash$^2$ \quad Shima Nofallah$^1$ \
Chintan Shah$^1$ \quad
Sai Chowdary Gullapally$^1$ \quad Limin Yu$^1$ \quad Michael Griffin$^1$ \quad Anand Sampat$^1$ \quad
 John Abel$^1$ \quad
Justin Lee$^1$ \quad 
Amaro Taylor-Weiner$^1$ \\
}

\institute{$^1$PathAI Inc, Boston, USA \quad $^2$Spring Discovery, California, USA  \\
\email{tan.nguyen@pathai.com}}

\maketitle              
\begin{abstract}
    Differences in staining and imaging procedures can cause significant color variations in histopathology images, leading to poor generalization when deploying deep-learning models trained from a different data source. Various color augmentation methods have been proposed to generate synthetic images during training to make models more robust, eliminating the need for stain normalization during test time. Many color augmentation methods leverage domain labels to generate synthetic images. This approach causes three significant challenges to scaling such a model. Firstly, incorporating data from a new domain into deep-learning models trained on existing domain labels is not straightforward. Secondly, dependency on domain labels prevents the use of pathology images without domain labels to improve model performance. Finally, implementation of these methods becomes complicated when multiple domain labels (e.g., patient identification, medical center, etc) are associated with a single image. We introduce ContriMix, a novel domain label free stain color augmentation method based on DRIT++, a style-transfer method. Contrimix leverages sample stain color variation within a training minibatch and random mixing to extract content and attribute information from pathology images. This information can be used by a trained ContriMix model to create synthetic images to improve the performance of existing classifiers. ContriMix outperforms competing methods on the Camelyon17-WILDS dataset. Its performance is consistent across different slides in the test set while being robust to the color variation from rare substances in pathology images. We make our code and trained ContriMix models available for research use.

\keywords{Synthetic data \and Domain Generalization \and Digital Pathology }
\end{abstract}
\section{Introduction}
Recent advancements in Machine Learning and slide digitization have transformed digital pathology by offering high-throughput, accurate analysis on large whole-slide images (WSI) \cite{MADABHUSHI2016170, bejnordi2017lymph}. However, pathology images often have large color variations across different labs and even within the same lab. This variation leads to poor performance of algorithms developed on certain domains (labs, scanners) when deployed on others.

Stain color normalization is often used to align the distribution of stain color of the test set to that of the training set. One way to do this is by extracting the color vectors of each stain from both sets, either from raw pixels \cite{ruifrok2001quantification}, using Singular Value Decomposition in Optical Density space \cite{macenko2009method}, or Non-negative Matrix Factorization \cite{vahadane2016structure}. Additionally, style-transfer methods have been proposed to perform stain normalization, leveraging frameworks such as \textit{pix2pix} \cite{salehi2020pix2pix}, StainGAN \cite{StainGAN}, StainNet \cite{StainNet}, and contrastive unpaired translation \cite{StainCUT}. 

Stain color augmentation is another method to address the generalization problem, which can lead to better performance than stain normalization \cite{tellez2019quantifying}. Augmentation generates several variations of input images with the same content but varied coloring to encourage the network to learn color-invariant features \cite{Tellez2019QuantifyingTE}. These methods can be divided into two groups. 

Most color augmentation methods in the first group rely on domain labels. For example, HistAuGAN \cite{wagner2021structure}, an application of DRIT++ to pathology, learns a one-to-many mapping based on disentangling the domain-invariant content (tissue morphology) in each image from the stain color attribute of each domain. Recently, Khamankar et al \cite{khamankar2023histopathological} suggested using adaptive instance normalization to create style-augmented synthetic images by mixing the style feature statistics of different images. These methods are dependent on domain labels and need to be retrained with every new domain, making scaling to new domains challenging. Additionally, these methods cannot take advantage of a large volume of unlabeled histopathology data to improve model performance. 

Without domain labels, one way to do color augmentation is to leverage the stain color vector extraction \cite{macenko2009method, vahadane2016structure} to extract stain vectors in Hematoxylin-Eosin (H\&E) images and use them to generate synthetic images by color transfer for training. Recently, deep-learning methods like STRAP \cite{yamashita2021learning} use style transfer to synthesize images with styles from medically irrelevant images while preserving the original high-level semantic content of pathology images.

We propose a novel color augmentation technique, ContriMix, an improvement over DRIT++ that does not require any domain labels. Like DRIT++, ContriMix disentangles the content of a pathology image (tissue morphology) from the stain color attributes (style). In contrast with DRIT++, ContriMix leverages the color difference between random pairs of samples from the training batch to train encoders for decoupling the content from the color attribute. Once trained, ContriMix can be used as a stain color augmentation technique to generate synthetic images to train other task-specific networks.

On Camelyon17-WILDS dataset, we demonstrate that backbone networks trained with ContriMix augmentation are capable of achieving color-invariant properties and outperform competing methods in a classification setting. Clustering of ContriMix representations shows that the content encodings are domain-invariant, while the attribute encodings capture color differences across different domains (hospitals). We further perform an in-depth subgroup analysis on slides from the test set and find that backbones trained with ContriMix augmentation have robust performance in presence of tissue patches containing a significant amount of red blood cells, lymphocytes, and low fractional tissue area.
We make the source code available for research use, along with ContriMix models trained on Camelyon17-WILDS and 2.5 million images from the Cancer Genome Atlas (TCGA) dataset.

\section{Method}
\subsection{Model architecture}
\begin{figure*}
    \centering
    \includegraphics[scale=0.30]{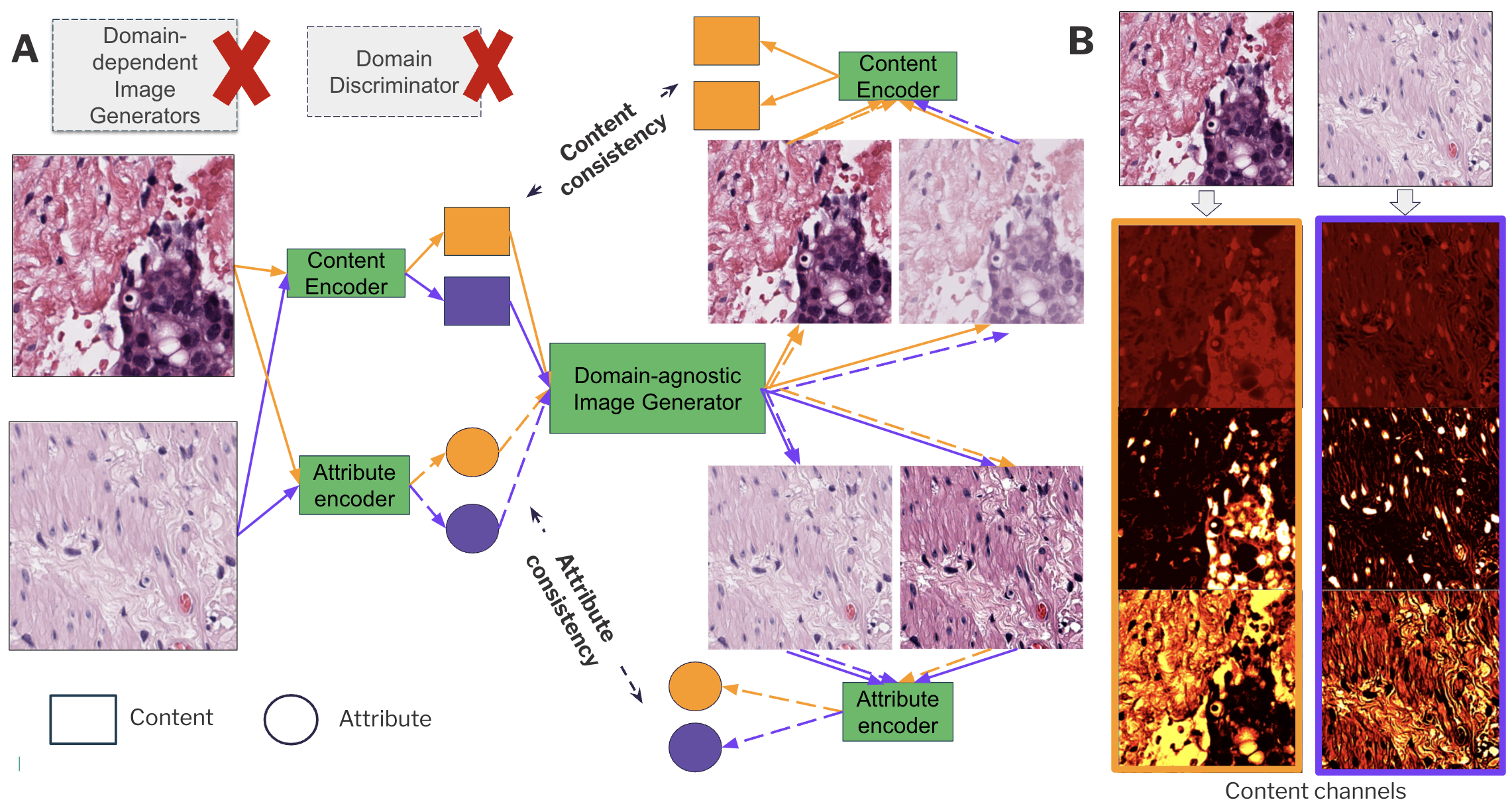}
    \caption{A) Overview of ContriMix - Content and attribute encodings are extracted, randomly mixed, and then combined to generate synthetic images without any domain labels. B) Three example content channels from input images. Different channels highlight different features in the tissue images.}
    \label{fig:contrimix}
\end{figure*}

Figure \ref{fig:contrimix}A shows the architecture of ContriMix. It consists of a content encoder \(E^c\) that extracts different tissue content such as cell nuclei, connecting tissue etc and an attribute encoder \(E^a\) that encodes the color appearance. It also includes an image generator \(G\) that takes a content encoding \(z^c\) (Figure \ref{fig:contrimix}B) and an attribute encoding \(z^a\) to generate a synthetic image. The image generator does not need the one-hot encoded domain to generate the output like DRIT++ \cite{lee2020drit++}.

In ContrixMix, all images from the training batch are passed to both encoders to extract the content encodings and attribute encodings. Next, randomly mixed combinations of the content and attribute encodings within the training minibatch are created and fed into the image generator \(G\) to create synthetic images. For simplicity, we will use \(I_{jk}\) to denote the synthetic image created from the content encoding \(z^c_j\) of the \(j^{th}\) image, \(I_j\), and attribute encoding \(z^a_k\) of \(I_k\), namely \(I_{jk} = G(z^c_j, z^a_k)\), \(z^c_j=E^c(I_j)\), and \(z^a_k=E^a(I_k)\).

The objective function of ContriMix is
\begin{equation}
    L_{ContriMix} = \lambda_a L_{attr.} + \lambda_c L_{cont.} + \lambda_{s} L_{self-recon.}, 
\end{equation}
where \(L_{attr.}, L_{cont.},\) and \(L_{self-recon.}\) are the attribute consistency loss, content consistency loss, and the self-reconstruction loss with respective weights \(\lambda_a, \lambda_c, \lambda_{s}\). Note that this objective function is much simpler than that of DRIT++ \cite{lee2020drit++} which requires adversarial losses for content and domain attribute encodings, latent space reconstruction loss, and KL divergence loss on the attribute encodings to enforce the attribute space to be distributed according to the standard normal distribution. 
Here, \(L_{attr.}\) measures the difference between the extracted attribute of the synthetic image \(E^a(I_{jk})\) and that of the self-reconstructed image \(E^a(I_{kk})\) to the expected attribute encoding \(z^a_k\). \(L_{cont.}\) measures the difference between the extracted content from the synthetic image \(E^c(I_{jk})\) and that of the self-reconstructed image \(E^c(I_{jj})\) to the expected content encoding  \(z^c_j\). Finally, \(L_{self-recon.}\) quantifies the difference between the self-reconstructed images \(I_{jj}\) and the input image \(I_j\). We use L1 losses in all of these loss terms.

\subsection{Competing methods for color augmentation}
We limit our comparison to other stain color augmentation methods, leaving out those that perform stain normalization to transfer the stain color of the test domain to the training domain. Competing methods include
\begin{itemize}
\item Reported methods from \href{https://wilds.stanford.edu/leaderboard/}{https://wilds.stanford.edu/leaderboard/}: ERM \cite{sagawa2021extending}, LISA \cite{yao2022improving}, IRMX (PAIR Opt) \cite{zhou2023pareto}, and ERM with targeted augmentation \cite{gao2022out}. 
\item HistAuGAN \cite{wagner2021structure}: We modified ERM training workflow to feed training images into the trained HistAuGAN networks to generate synthetic images with a probability of 0.5. HisAuGAN networks were frozen when training the backbone. The weights of HistAuGAN networks are provided by \cite{wagner2021structure}. 
For attribute sampling, we explored two options 1) Sampling with attributes from 3 training domains only and 2) Sampling with attributes from all 5 domains. 
The first one mimics a practical scenario where no data from the validation set and the test set are available during training, while better performance is expected in the second. We note that the HistAuGAN model from \cite{wagner2021structure} was trained on all 5 domains using whole-slide images from Camelyon-17, giving it an advantage in terms of data diversity compared to ContriMix. For each option, we trained the backbone for 40 epochs using the AdamW \cite{loshchilov2017decoupled}  optimizer with a learning rate of 1e-4 for 5 random seeds.
\item Recent methods: STRAP \cite{yamashita2021learning}  and FuseStyle \cite{khamankar2023histopathological}. Due to the lack of publicly available trained models train for Camelyon17-WILDS (STRAP) or source code (FuseStyle), we were unable to compare validation accuracy . The test accuracy comparison is provided in Table \ref{camelyon17-results-table}.
\end{itemize}

\section{Results and Discussion}
\subsection{Dataset}
The Camelyon17-WILDS dataset contains 450,000 H\&E stained 96 x 96 image patches from 5 hospitals. The objective is to classify them to either tumor or normal. The training dataset consists of patches from the first 3 hospitals, while the validation and test datasets are from the 4\textsuperscript{th} and 5\textsuperscript{th} hospitals, see Fig. \ref{fig:performance} A
\subsection{ContriMix training}
To evaluate the effect of ContriMix, we train a DenseNet121 backbone from scratch on the binary classification task using the training split and compared the performance on the out-of-domain test split following the protocol in \cite{koh2021wilds}. We modified the baseline ERM workflow to insert the two ContriMix encoders and image generator between the input and the backbone. We used a weighted sum of the binary cross-entropy loss and the component losses from ContriMix, 
\begin{equation}
        L_{total} =  \lambda_{BCE} L_{BCE} + \lambda_{s} L_{self-recon.} + \lambda_a L_{attr.} + \lambda_c L_{cont.} 
\end{equation}
where \(\lambda_{BCE}\)=0.5, \(\lambda_{s}\)=0.1, \(\lambda_{a}\)=0.1, \(\lambda_{c}\)=0.3. We explored different combinations of weights and found that they mainly impact the speed of convergence and not the backbone performance. Moreover, training ContriMix networks separately or jointly with the backbone yielded no significant difference in the backbone performance while joint training was more convenient and slightly faster. Therefore, we used joint training. We used the AdamW \cite{loshchilov2017decoupled}  optimizer with a learning rate of 1e-4 and an L2-regularization of 1e-4 for all 10 random seeds. Input images are randomly rotated by multiples of 90 degrees, randomly flipped, and passed to ContriMix encoders. The training time was 12 GPU-hours (RTX8000).

\subsection{Benchmarking  results}
Table \ref{camelyon17-results-table} reports the performance of DenseNet121 backbones trained with different color augmentation methods. ContriMix outperformed other methods in terms of average accuracy on the test set while being second to the ERM with targeted augmentation on the validation set. The test accuracy of ContriMix augmentation is significantly higher than that of HistAuGAN (3-domains augmentation) by 23.2\% while being trained on the same data. Interestingly, ContriMix augmentation trained on 3 hospitals also surpasses other augmentation methods trained from data-abundant sources such as HistAuGAN 5-domains and STRAP. 

\begin{table}
  \caption{\textbf{Performance comparison on Camelyon17-WILDS}.}
  \label{camelyon17-results-table}
  \centering
  \begin{tabular}{lll}
    \toprule
    \textbf{\# Method}& \textbf{OOD Val Acc. (\%)} & \textbf{Test Acc. (\%)} \\
    \midrule
    ERM (rand search) \cite{sagawa2021extending} &  85.8 $\pm$ 1.9 & 70.8 $\pm$ 7.2    \\
    HistAuGAN (3 dom.) \cite{wagner2021structure} & 85.8 $\pm$ 1.1 & 71.4 $\pm$ 7.4 \\
    IRMX (PAIR Opt) \cite{zhou2023pareto} &  84.3 $\pm$ 1.6 & 74.0 $\pm$ 7.2    \\
    LISA \cite{yao2022improving} &  81.8 $\pm$ 1.4 & 77.1 $\pm$ 6.9    \\
    FuseStyle \cite{khamankar2023histopathological} & - & 90.49 (-)\\
    ERM w/ targeted aug \cite{gao2022out}&  \textbf{92.7 $\pm$ 0.7} & 92.1 $\pm$ 3.1    \\
    HistAuGAN (5 dom.) \cite{wagner2021structure} & 87.9 $\pm$ 2.3 & 92.6 $\pm$ 0.7 \\
    STRAP \cite{yamashita2021learning} & - & 93.7 $\pm$ 0.15 \\
    \textbf{ContriMix} &  \textbf{91.9 $\pm$ 0.6} & \textbf{94.6 $\pm$ 1.2}    \\
    \bottomrule
  \end{tabular}
\end{table}

\begin{table}
  \caption{\textbf{Ablation experiments for number of training centers}. We study the impact of dropping entire domains on ContriMix.}
  \label{num-centers-table}
  \centering
  \begin{tabular}{c|c|c}
    \toprule
    \textbf{\# Train Centers}& \textbf{OOD Val Acc.(\%)} & \textbf{Test Acc.(\%)} \\
    \midrule
    3 &  91.9 $\pm$ 0.6 & 94.6 $\pm$ 1.2    \\
    2 &  87.2 $\pm$ 1.3 & 88.8 $\pm$ 1.8    \\
    1 &  85.6 $\pm$ 1.4 & 86.9 $\pm$ 4.0    \\
    \bottomrule
  \end{tabular}
\end{table}

\begin{figure*}
    \centering
    \includegraphics[scale=0.36]{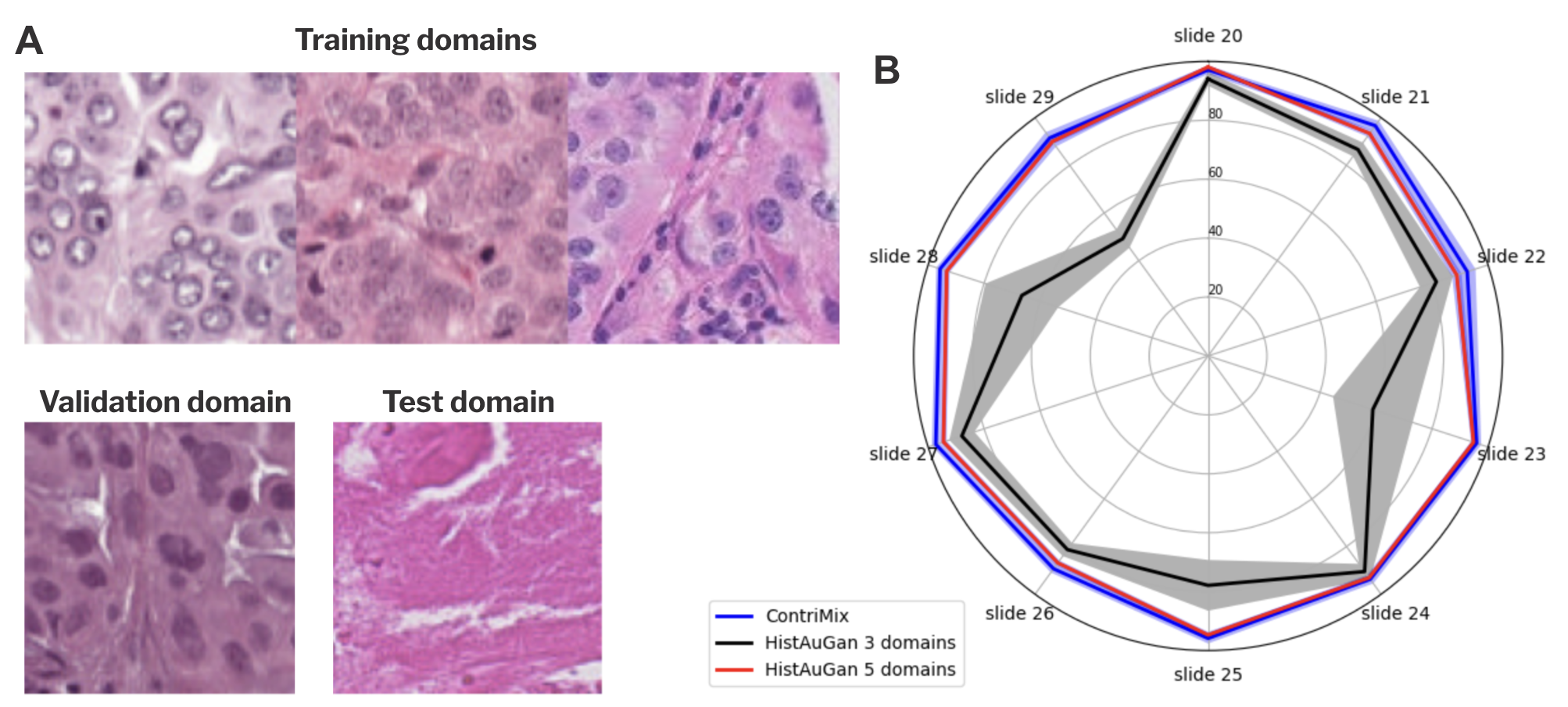}
    \caption{A) Histopathology images from different hospitals in Camelyon-17-WILDS exhibit significant color variation. B) Performance comparison of DenseNet121 backbones trained with ContriMix augmentation, HistAuGAN 3-domains and 5-domains augmentation on 10 different test slides.}
    \label{fig:performance}
\end{figure*}

\begin{figure*}
    \centering
    \includegraphics[scale=0.35]{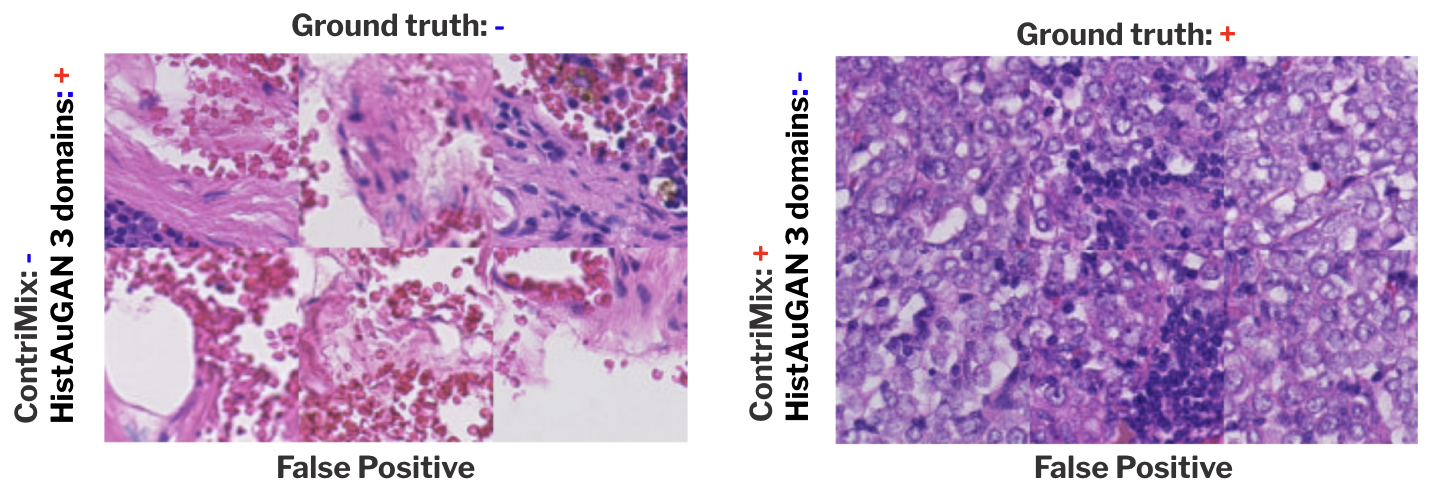}
    \caption{Six false positive patches (left) and six false negative patches (right) from HistAuGAN 3-domains correctly classified by ContriMix}
    \label{fig:misclassified}
\end{figure*}

 Figure \ref{fig:performance}B compares the test accuracy of ContriMix against HistAuGAN 3-domains and 5-domains at a WSI level, with error bars denoting the \(\pm 1\) standard deviation from mean accuracy. 
 We observed significant performance gaps of mean accuracy on slides 23 (37.1\%), 28 (29.2\%), and 29 (42.3\%) between ContriMix and HistAuGAN 3-domains while both are trained using data from the same domains.
Upon further inspection, we discovered that the observed gaps were due to the presence of patches with significant amount of red blood cells, patches that are located near tissue margin, or a high number of lymphocytes with dark stained nuclei causing significant color variation (see Figure \ref{fig:misclassified} for examples).

To our knowledge, the highest reported test accuracy is 97.5\% from ContextViT \cite{bao2023contextual}. Due to the use of a different backbone (ViT-S/8) with 2.75\(\times\) more parameters than DenseNet121 and the use of an additional 1.8 million unlabeled Camelyon17-WILDS patches for pre-training, we did not include ContextViT in our comparison. 

\begin{figure*}
    \centering
    \includegraphics[scale=0.2]{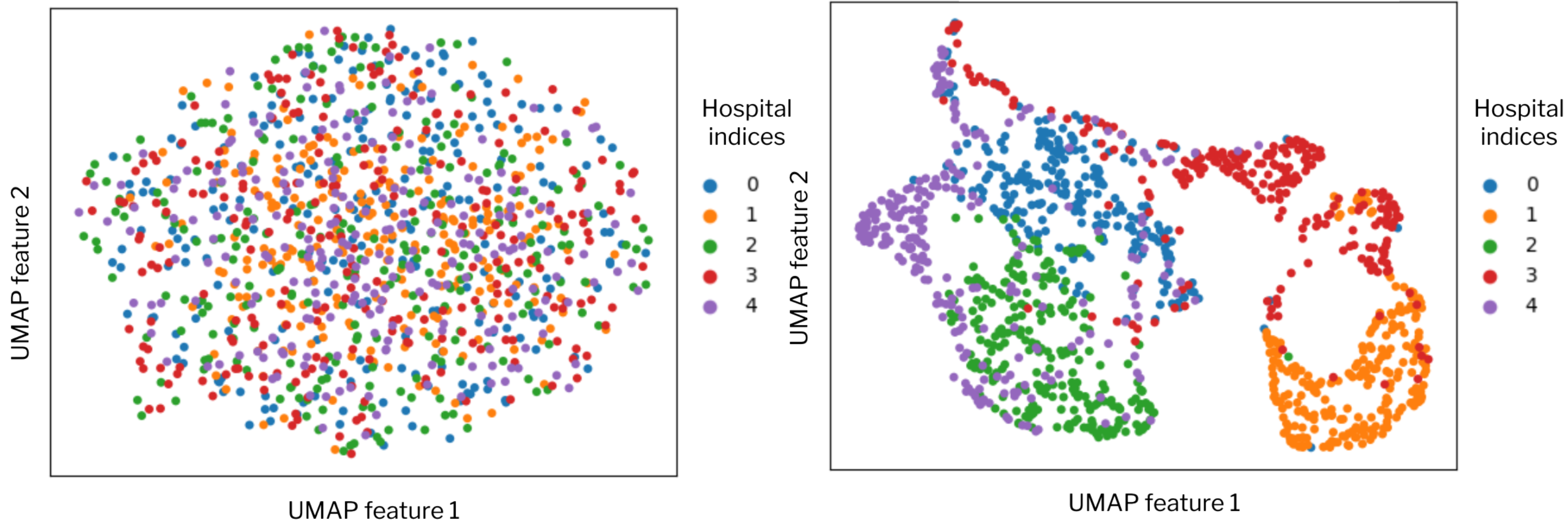}
    \caption{UMAP plot for ContriMix content (left) and attribute (right) encodings colored by different centers in Camelyon17-WILDS.}
    \label{fig:ContriMix-umap}
\end{figure*}

To visualize the dependency between the encodings from ContriMix and domains, we pass encodings of 7200 patches to UMAP \cite{mcinnes2018umap}. Figure \ref{fig:ContriMix-umap} shows that the attribute tensors contain the differences across patches from different domains, while the content encoder learns center-invariant features. \textit{This happens even though there is no access to domain supervision during training.}

\subsection{Ablation study - Diversity of training domains}
In this ablation, we remove data belonging to different training domains and study its impact on ContriMix. This serves to simulate the real-world setting where we are starved of domain-diverse data. We choose to keep the centers with the least number of samples in the train set - for training with one center, we keep only center 0, while for training with two centers, we keep centers 0 and 3.
While there is a drop in performance (Table \ref{num-centers-table}), ContriMix with one center is still able to outperform several methods trained on 3 centers, as seen in Table \ref{camelyon17-results-table}. The results indicate that ContriMix is better able to utilize the intra-dataset variations even in the presence of a single domain. 

\subsection{Qualitative evaluation by a board-certified pathologist}
We conducted an expert evaluation of the synthetic images generated by ContriMix (Supplementary section). Eighty patches were randomly selected and shared with a board-certified pathologist, with the following question -
`Please evaluate the quality of the synthetic images. Please label the quality as 'NOT SATISFACTORY' if the synthetic image includes any artifact that was not present in the original image, or changes any biological details in the original image'. The pathologist's feedback is as follows - \textit{`All the synthetic images are free of artifacts or changes that would hinder pathologic interpretation'}. 

The supplementary section additionally contains examples of ContriMix's robustness to image artifacts (ink, blur, tissue folding from processing), ablations around mixing parameters and pseudocode. 

\section{ContriMix for Multiple Instance Learning}
\begin{figure*}
  \centering
    \includegraphics[scale=0.5]{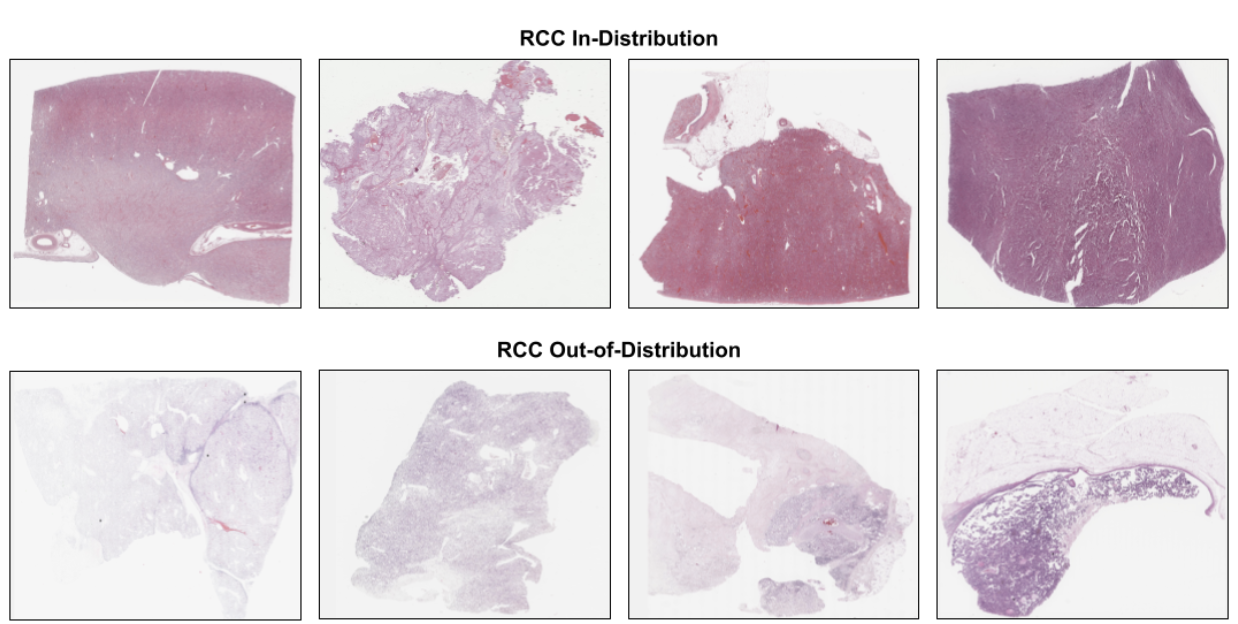}
    
    \caption{ Examples of in-distribution (ID) and out-of-distribution (OOD) WSIs from the RCC cancer subtyping datasets. The first row shows ID TCGA WSIs, while the second row shows WSIs procured from a different lab site and scanner. We can see the variations in tissue preparation and scanning which lead to significant visual differences.}

    \label{fig:id-and-ood}
\end{figure*}

To demonstrate ContriMix color augmentation for WSI-level prediction, we study the problem of prediction of cancer subtypes in renal cell carcinoma (RCC) within The Cancer Genome Atlas (TCGA) \cite{weinstein2013cancer}. TCGA-RCC has 948 WSIs with three histologic subtypes: 158 WSIs with chromophobe RCC (KICH), 504 WSIs belonging to clear cell RCC (KIRC), and 286 to papillary RCC (KIRP). For the OOD dataset, we use WSIs acquired from an independent laboratory. These WSIs have been acquired with different image processing and laboratory techniques, leading to notable differences in staining appearance and visual characteristics. Some example images highlighting this contrast with TCGA can be found in Fig. \ref{fig:id-and-ood}. The class-wise distribution for the OOD dataset is as follows: 46 KICH, 254 KIRC, and 134 KIRP WSIs. 

For RCC cancer subtype prediction, we use the Deep Attention Multiple Instance Learning (AB-MIL) \cite{ilse2018attention} as a baseline. Synthetic patches are generated via ContriMix, with the same architectures for encoders and generators as for the Camelyon17-WILDS, and passed along with the original patches to the MIL model. For the same reason with Camelyon17-WILDS, we restrict our comparisons to stain color augmentation methods. Other competing methods include the stain color augmentation based on stain vector extraction using the Macenko method \cite{macenko} (referred to as `StainAug' in Table \ref{rcc-subtype-table}), and Supervised Contrastive MIL (SC-MIL) \cite{juyal2023scmil}, which has been shown to improve OOD performance on MIL datasets. All MIL models were trained on non-overlapping patches of size 224 x 224 pixels at a resolution of 1 micron per pixel, with a total of 800k patches extracted approximately. Bag sizes were varied from 24 to 1500, and batch sizes were between 8 and 32. An Imagenet-pretrained ShuffleNet was used as the feature extractor. Augmentations applied included flips, center crops, HSV transforms, and probabilistic grayscaling; the values were kept consistent across all techniques. SC-MIL training was done with a temperature value of 1.0 \cite{juyal2023scmil}. An Adam optimizer with a learning rate of 10\textsuperscript{-4} was used for all models.

Exhaustive sampling of patches was done at inference time, and majority vote across bags was used as the WSI level prediction to compute the macro-average F1 score and macro-average 1-vs-rest AUROC. While SC-MIL has the best performance on the ID test set with AB-MIL with ContriMix stands second. AB-MIL with ContriMix has the best performance on the OOD test set, highlighting the effectiveness of using ContriMix in the MIL framework.

\begin{center}
\begin{table*}
  \caption{\textbf{Comparison of predictive performance on RCC subtyping.} Mean \(\pm\) standard deviation were computed via bootstrapping over 1000 runs on the test set.} 
  \label{rcc-subtype-table}
  \centering
  \begin{tabular}{l|cc|cc}
    \toprule
    \textbf{\# Method}& \textbf{ID Test F1} & \textbf{ID Test AUROC} & \textbf{OOD Test F1} & \textbf{OOD AUROC} \\
    \midrule

AB-MIL             & 87.80 \(\pm\) 2.29  & 97.78 \(\pm\) 0.61  & 74.85 \(\pm\) 2.42  & 91.75 \(\pm\) 1.61   \\
AB-MIL+StainAug  & 87.46 \(\pm\) 2.28   & 97.83 \(\pm\) 0.68   & 72.83 \(\pm\) 2.34    & 94.59 \(\pm\) 1.19    \\
SC-MIL             & \textbf{90.91 \(\pm\) 1.98}    & \textbf{98.07 \(\pm\) 0.40}   & 77.83 \(\pm\) 2.39   & 95.03 \(\pm\) 1.09  \\
AB-MIL+ContriMix    & 89.10 \(\pm\) 2.12    & 97.69 \(\pm\) 0.70  & \textbf{81.23 \(\pm\) 2.32}  & \textbf{95.59 \(\pm\) 1.27}   \\
    \bottomrule
  \end{tabular}
\end{table*}
\end{center}

\section{Limitations}
At present, there is no systematic way to determine the optimum number of content and attribute channels for ContriMix. A larger than necessary number of attributes may lead to an encoding of redundant information, longer training time but marginal gains in terms of representing true data diversity. In our experiments, a simple hyper-parameter search sufficed, however running this on other image modalities like immunohistochemistry may yield different results. 

\section{Conclusion}
In summary, we introduce ContriMix, a scalable technique for stain color augmentation for histopathology images. Through simple mixing of content and attribute within training minibatch along with consistency-based losses, ContriMix can synthesize realistic images with different color appearances while preserving tissue morphology. ContriMix does not require any information about the domain of training patches. This key advantage allows using a trained ContriMix model to extract the stain color (style) from a vast body of unlabeled images and use them to further increase the diversity of synthetic images for color augmentation. We demonstrate that backbones trained with ContriMix color augmentation have a better out-of-domain accuracy compared to other color augmentation methods on Camelyon WILDS-17. Our ablation studies suggest the effectiveness of ContriMix for generating domain-invariant representations without needing domain labels, along with desirable properties such as robustness to color variation from rare substances and learning meaningful representations even in data-diversity starved regimes. We release our code and trained models for research use.

\newpage
\bibliographystyle{splncs04}

\bibliography{main}

\setcounter{section}{0}

\title{Supplementary Section for ContriMix: Scalable stain color augmentation for domain generalization without domain labels in digital pathology}
\author{}
\institute{}
\maketitle         

\section{Additional ablation studies}
\subsection{Number of mixes}
Here, we investigate different numbers of mixes which the combination of content and attribute tensors to get a synthetic image with the same content, ranging from 1 to 5, with all other hyperparameters fixed. The results in Table \ref{num-mixing-table} indicate that increasing the number of mixes beyond a certain limit (in this case, 4) on Camelyon17-WILDS dataset has no significant effect on the model performance. This finding enables us to use a lower number of mixes and larger minibatch size at training time.

\begin{table}
  \caption{\textbf{Ablation experiments for the number of mixes}. Mean \(\pm\) standard deviation accuracies from 10 random seeds are reported.}
  \label{num-mixing-table}
  \centering
  \begin{tabular}{lll}
    \toprule
    \textbf{\# Mixes}& \textbf{OOD Val Acc. (\%)} & \textbf{Test Acc. (\%)} \\
    \midrule
    1  &  92.0 \(\pm\) 0.7  &  92.4 \(\pm\) 3.0 \\ 
    2  &  92.2 \(\pm\) 0.9  &  90.8 \(\pm\) 6.1 \\ 
    3  &  91.8 \(\pm\) 1.1  &  93.9 \(\pm\) 1.7 \\ 
    4  &  91.9 \(\pm\) 0.6  &  94.6 \(\pm\) 1.2 \\ 
    5  &  92.4 \(\pm\) 0.8  &  93.2 \(\pm\) 2.3 \\ 
    \bottomrule
  \end{tabular}
\end{table}

\subsection{Random vs. targeted mixing}
ContriMix augmentation uses a default method of attribute selection that involves randomly selecting attributes from a minibatch to combine with an image's content. This mixing algorithm does not take into account the domain identifiers when selecting attributes. We ran an experiment to investigate the effect of targeted mixing versus random mixing. In targeted mixing, the domain identifiers of the attributes and content are mutually exclusive. For example, if the content is from domain 1, the attribute can only be chosen from images in domain 2 or 3. 
Table \ref {mixing-methods-table} shows the comparison of random vs targeted mixing on the Camelyon17-WILDS dataset. The number of mixes for this experiment was 4. The experiments are run on 10 random seeds. All other hyper-parameters are the same. Even though there exists a domain imbalance ratio of approximately 1:3 across the training domains, unsupervised random mixing still performs on par with targeted mixing. 

\begin{table}
  \caption{\textbf{Ablation experiments for random vs. targeted mixing}. Mean \(\pm\) standard deviation accuracies from 10 random seeds are reported.}
  \label{mixing-methods-table}
  \centering
  \begin{tabular}{lll}
    \toprule
    \textbf{Mixing method} & \textbf{OOD Val Acc.(\%)} & \textbf{Test Acc.(\%)} \\
    \midrule
    Random mixing &  91.9  \(\pm\) 0.6  &  94.6 \(\pm\) 1.2   \\
    Targeted mixing &  91.9 \(\pm\) 0.8 & 93.7 \(\pm\) 1.3  \\

    \bottomrule
  \end{tabular}
\end{table}

\subsection{Number of attributes}
We vary the number of attributes to study the effect of additional representational capacity in the ContriMix. The results are reported in Table \ref{num-attributes-table}. Increasing the number of attributes helps the model learn until a certain point, beyond which performance starts to saturate.

\begin{table}
  \caption{\textbf{Ablation experiments for varying the number of attributes }. Mean \(\pm\) standard deviation accuracies from 10 random seeds are reported.}
  \label{num-attributes-table}
  \centering
  \begin{tabular}{lll}
    \toprule
    \textbf{\# Attributes} & \textbf{OOD Val Acc.(\%)} & \textbf{Test Acc.(\%)} \\
    \midrule
    3 &  92.1 \(\pm\) 1.1 & 92.8 \(\pm\) 2.2    \\
    5 &  92.4 \(\pm\) 0.9 & 93.8 \(\pm\) 1.1    \\
    7 &  91.9 \(\pm\) 0.6 & 93.1 \(\pm\) 0.9    \\
    9 &  92.7 \(\pm\) 1.0 & 94.1 \(\pm\) 1.4    \\
    11 &  92.5 \(\pm\) 0.8 & 93.4 \(\pm\) 2.5    \\
    13 &  92.0 \(\pm\) 1.3 & 94.1 \(\pm\) 1.3    \\
    \bottomrule
  \end{tabular}
\end{table}

\begin{figure}
    \centering
    \includegraphics[width=8cm]{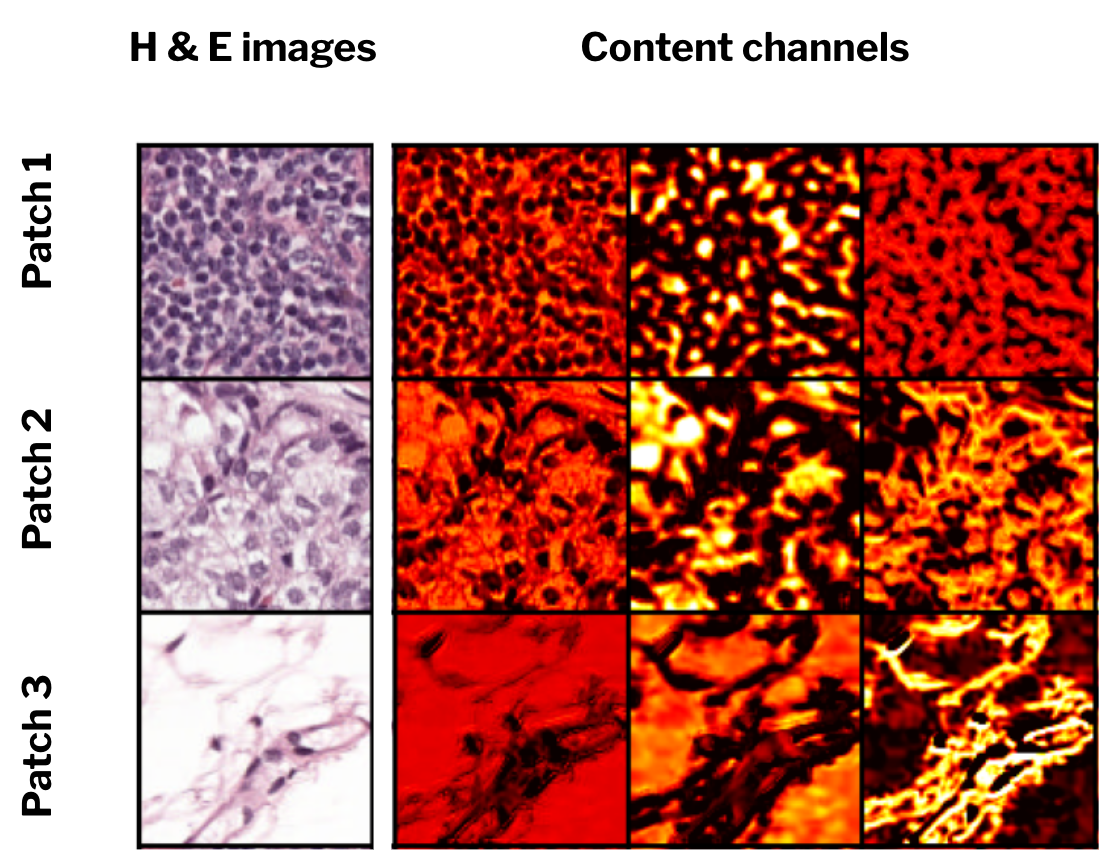}
    \caption{Content channels of ContriMix for three different input images. The left most column contains the original images. The next three columns show three different content channels. ContriMix learns to encode biological information in different channels. Table \ref{content-channels-table} outlines the biological details in the channels.}
    \label{fig:ContriMix-content}
\end{figure}

\subsection{Content tensors}
\label{content-vis}
Another qualitative study is conducted to gain insights into what the content channels are learning by visualizing the content channels for 100 patches. These content maps were shared with a board-certified pathologist, who was asked to determine whether any biological details were being encoded in each channel. If so, the pathologist was asked to identify what concept(s) they correspond to. An example is shown in Fig. \ref{fig:ContriMix-content}, which has 3 rows, one for each image. The leftmost column shows the original images. The next 3 columns show the extracted content maps. According to the evaluation, different content channels appear to learn various kinds of details, which are tabulated in Table \ref{content-channels-table}. We emphasize that no annotations were used in training to teach the model to identify these structures specifically.

\begin{table}
  \caption{Expert evaluation of ContriMix content channels and the biological details captured by them.}
  \label{content-channels-table}
  \centering
  \begin{tabular}{lp{0.6\columnwidth}}
    \toprule
    \textbf{Content channel} & \textbf{Details}  \\
    \midrule
    1 &  Cytoplasm, background, connecting tissue   \\
    2 &  Acellular area (Lumen, blood vessel, background)     \\
    3 & Nuclei, adipose tissue    \\
    \bottomrule
  \end{tabular}
\end{table}

\section{The image generator of ContriMix}
In all of our experiments, we use a dot product for the image generator \(G\). This is inspired by the physics of histochemistry image formation. Following the derivation in \cite{ruifrok2001quantification}, the optical density can be written as \(OD = -log(I / I_o) = CM\). Here, \(C\) is a concentration matrix with each row containing the stain concentration at each pixel. \(M\) is a stain color vector matrix where rows are the color vectors. \(I\) is the raw intensity image obtained from the camera, \(I_o\) is the background intensity. One can associate content tensor \(z^c\) and the attribute tensor \(z^a\) extracted by ContriMix  with the stain concentration \(C\) and the stain vector matrix \(M\) respectively in the optical density equation. Moreover, this model also suggests that a tensor dot product can be used for the image generator \(G(z^c, z^a) = z^c \cdot z^a\). 

\section{Example synthetic images generated by ContriMix}
\begin{figure*}
    \centering
        \centering
        \includegraphics[width=\textwidth]{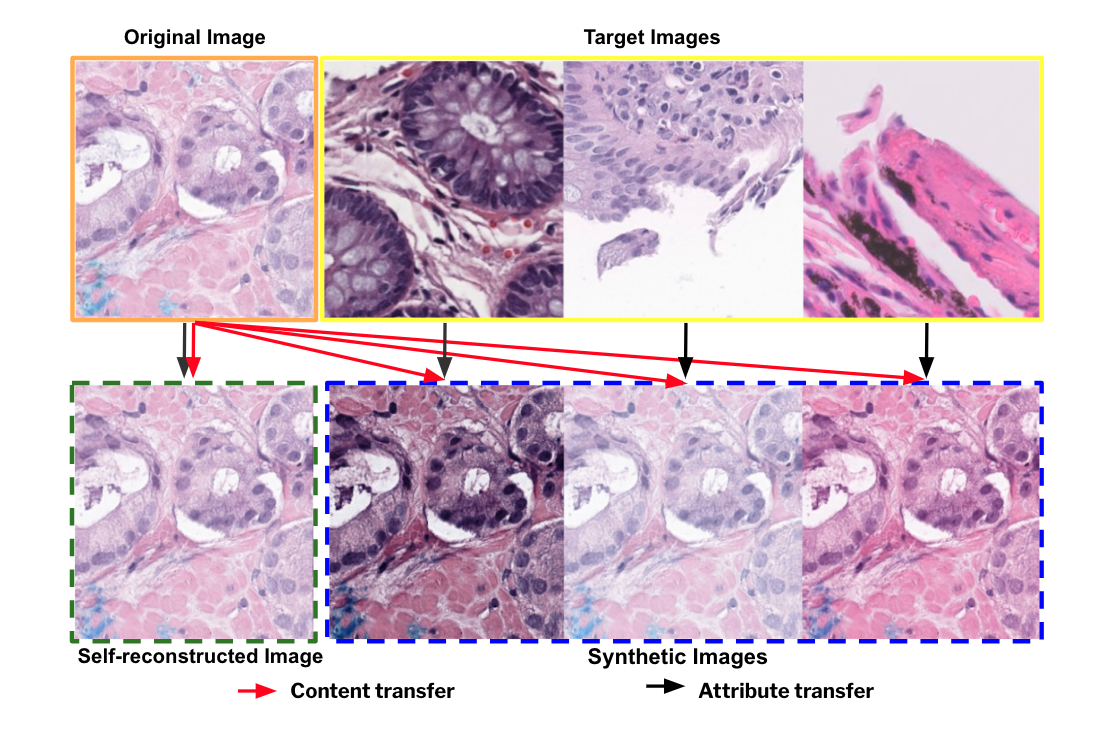}
        \label{fig:ContriMix-s1}
    \hfill
        \centering
        \includegraphics[width=\textwidth]{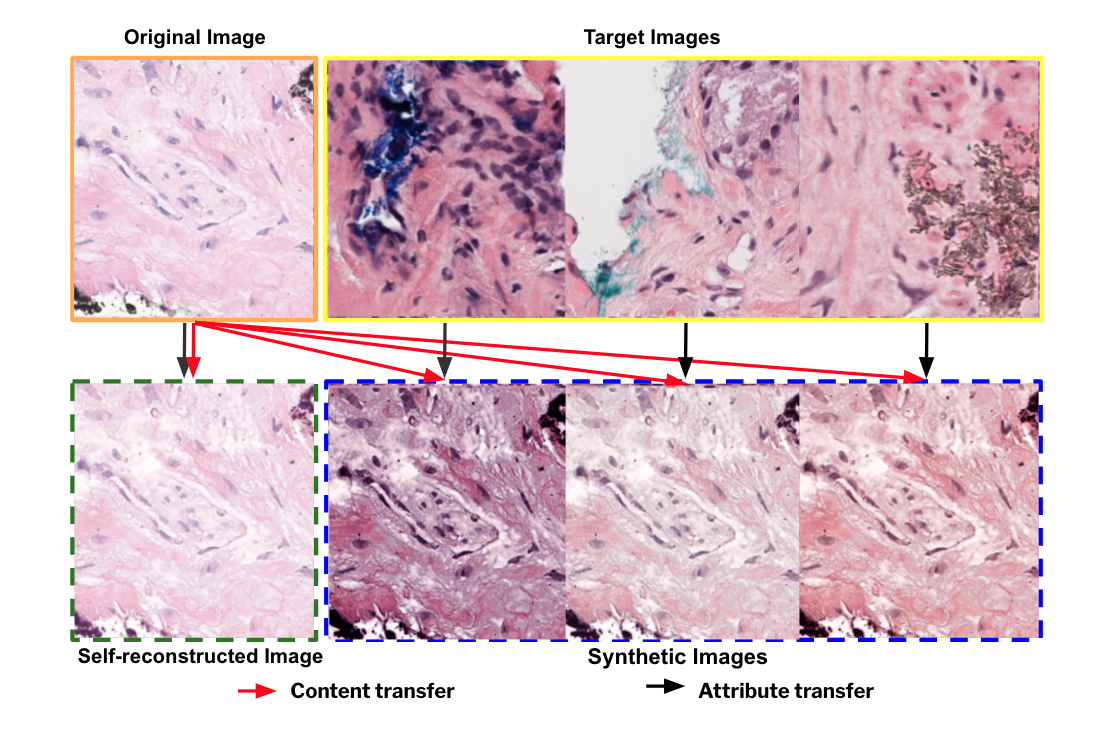}
        \label{fig:ContriMix-s2}
\end{figure*}

\begin{figure*}
    \hfill
        \centering
        \includegraphics[width=\textwidth]{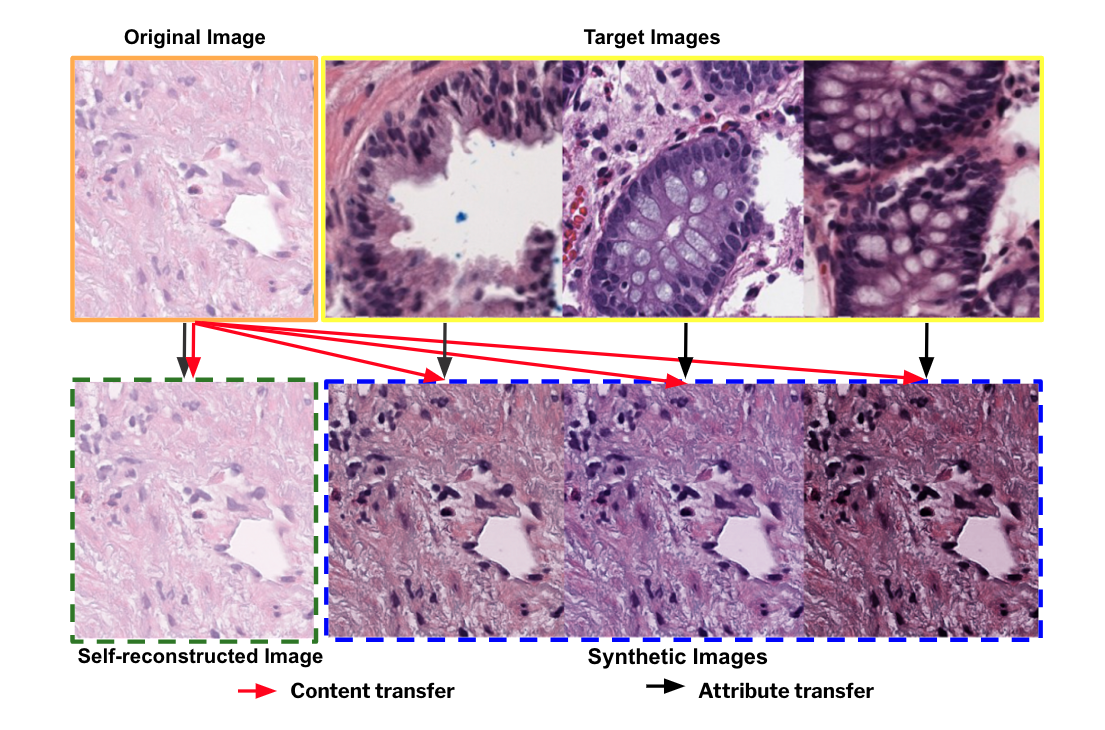}
        \label{fig:ContriMix-s3}
    \hfill
        \centering
        \includegraphics[width=\textwidth]{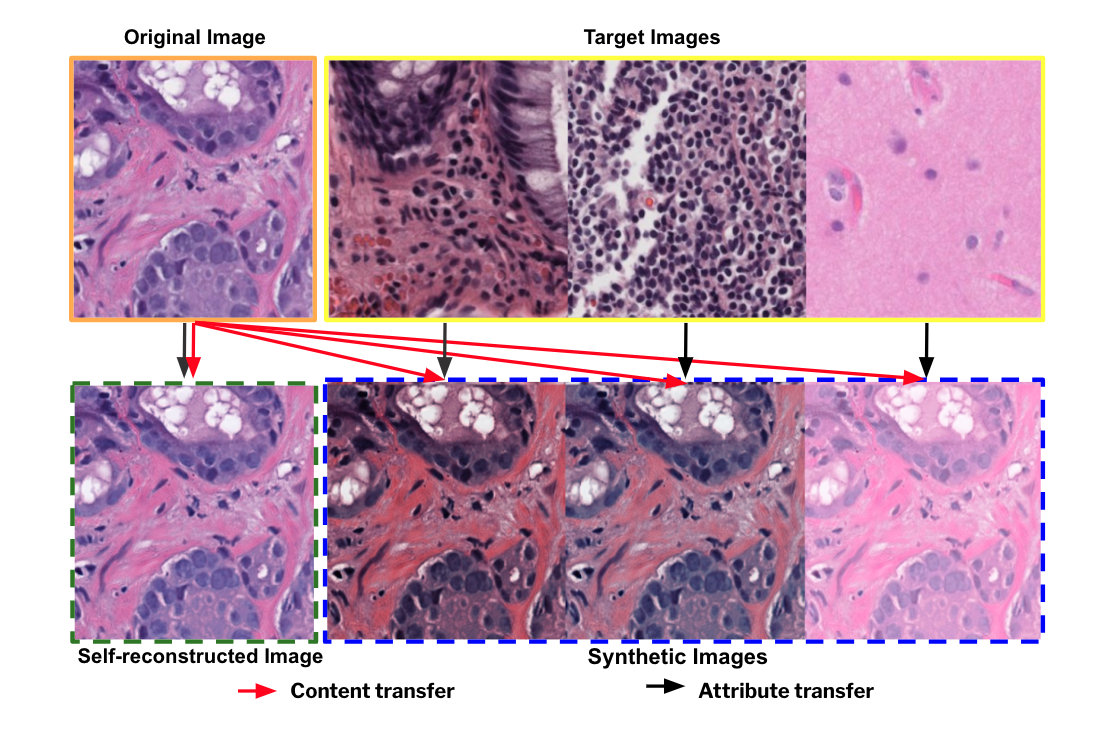}
        \label{fig:ContriMix-s4}
\end{figure*}

\setcounter{figure}{2}
\begin{figure*}
    \centering
    \includegraphics[width=\textwidth]{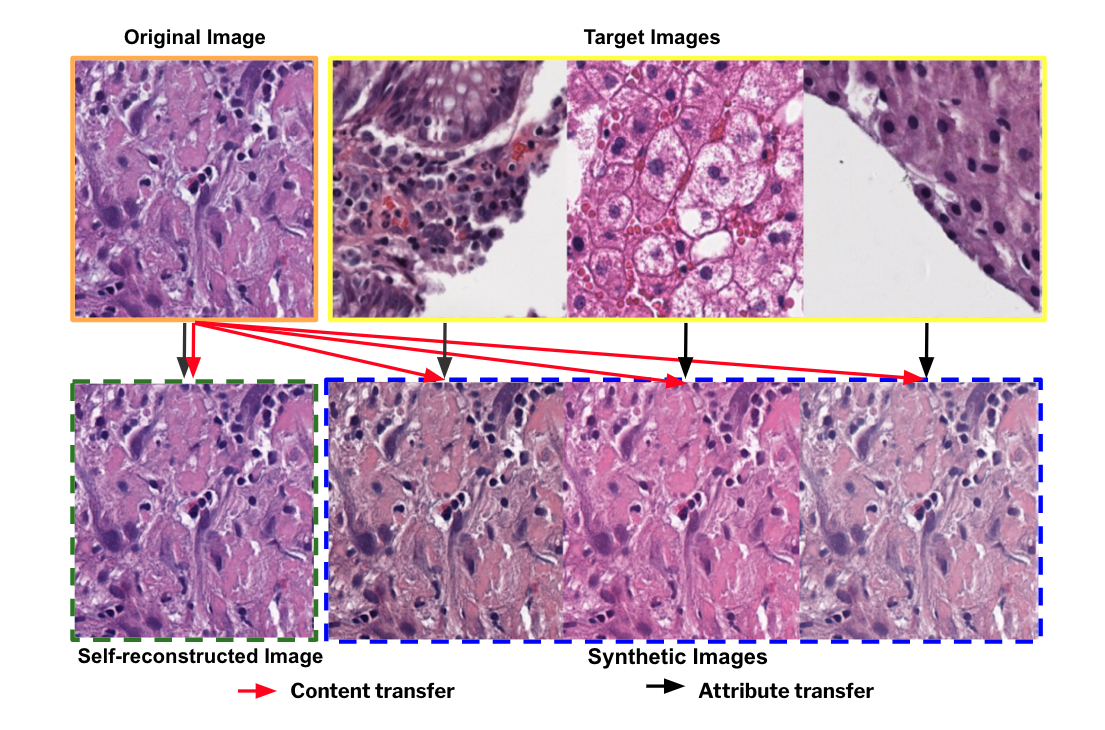}
    \label{fig:ContriMix-s5}
    \caption{Different examples of synthetic images generated by ContriMix. ContriMix attribute tensors learn to ignore artifacts (e.g.- marker ink, black spots) while the content tensors preserve relevant information without introducing hallucinations. Apart from artifacts, ContriMix is able to account for the presence of background pixels in the input images.}
\end{figure*}

\clearpage
\section{ContriMix pseudocode (PyTorch-style)}
\begin{verbatim}
# N: Size of the minibatch
# L: Number of attributes
# M: Number of mixings per image
# E_c: Content encoder
# E_a: Attribute encoder 
# G: Image generator
# lambda_s: Self-reconstruction loss weight
# lambda_a: Attribute consistency loss weight
# lambda_c: Content consistency loss weight


# Load a batch with N samples
for b in loader: 
    target_idxs = torch.randint(0, N, size=(N, M))
    zc = E_c(b)
    za = E_a(b)

    l1_loss = torch.nn.L1Loss()
    b_sr = G(zc, za)  #Batch self-reconstruction

    sr_loss = l1_loss(b_sr, b) #Self-reconstruction loss

    attr_cons_losses = [l1_loss(E_a(b_sr), z_a)]  #Attribute consistency losses 
    
    cont_cons_losses = [l1_loss(E_c(b_sr), z_c)]  #Content consistency losses 
    
    for mix_idx in range(M):
        za_tgt = za[target_idxs[:, mix_idx]]  #Target attribute for mixing
        
        b_ct = G(zc, za_target)  # Synthetic image
        attr_cons_losses.append(l1_loss(E_a(b_ct), za_target))
        cont_cons_losses.append(l1_loss(E_c(b_ct), zc))

    # Avergage over the mixing dimension
    attr_loss = torch.mean(torch.stack(attr_cons_losses, dim=0))
    cont_loss = torch.mean(torch.stack(cont_cons_losses, dim=0))


    # Loss
    loss = lambda_s * self_recon_loss + lambda_a * attr_loss + lambda_c * cont_loss
    
    # Optimization step
    loss.backward()
    optimizer.step()
\end{verbatim}
\clearpage

\end{document}